%% file: main.tex
\documentclass[aps, prmaterials, reprint, amsfonts, amssymb, amsmath, nofootinbib, twoside, superscriptaddress]{revtex4-2}
\usepackage[english]{babel}
\usepackage[utf8]{inputenc}
\usepackage[colorinlistoftodos, color=green!40, prependcaption]{todonotes}
\input{preamble}
\usepackage[pdftex, pdftitle={Towards Experimentally Realizable Superconductors with Disorder-Aware Guided Diffusion}, pdfauthor={Andrew F. Valentini et al.}]{hyperref}
\usepackage{comment}
\usepackage[normalem]{ulem}
\usepackage{xcolor}
\usepackage{booktabs}
\usepackage{threeparttable}
\usepackage{makecell}
\usepackage{chemformula}

\begin{document}
\title{Towards Experimentally Realizable Superconductors with Disorder-Aware Guided Diffusion}

\author{Andrew F. Valentini}
    \affiliation{Department of Physics, University of Florida, Gainesville, Florida 32611, USA}
    \affiliation{Department of Materials Science and Engineering, University of Florida, Gainesville, Florida 32611, USA}
    \affiliation{Quantum Theory Project, University of Florida, Gainesville, Florida 32611, USA}

\author{Gabriele D. Di Gianluca}
    \affiliation{Department of Physics, University of Florida, Gainesville, Florida 32611, USA}

\author{Pawan Prakash}
    \affiliation{Department of Physics, University of Florida, Gainesville, Florida 32611, USA}
     \affiliation{Department of Materials Science and Engineering, University of Florida, Gainesville, Florida 32611, USA}
    \affiliation{Quantum Theory Project, University of Florida, Gainesville, Florida 32611, USA}

\author{James Grossmann}
    \affiliation{Department of Physics, University of Florida, Gainesville, Florida 32611, USA}
   
\author{Jung Soo Kim}
    \affiliation{Department of Physics, University of Florida, Gainesville, Florida 32611, USA}
    
\author{Melchior Jean}
    \affiliation{Department of Physics, University of Florida, Gainesville, Florida 32611, USA}
    
\author{Benjamin Geisler}
    \affiliation{Department of Physics, University of Florida, Gainesville, Florida 32611, USA}
    \affiliation{Department of Materials Science and Engineering, University of Florida, Gainesville, Florida 32611, USA}
    \affiliation{Quantum Theory Project, University of Florida, Gainesville, Florida 32611, USA}

\author{Gregory R. Stewart}
    \affiliation{Department of Physics, University of Florida, Gainesville, Florida 32611, USA}

\author{James J. Hamlin}
    \affiliation{Department of Physics, University of Florida, Gainesville, Florida 32611, USA}
  
\author{Peter J. Hirschfeld}
    \affiliation{Department of Physics, University of Florida, Gainesville, Florida 32611, USA}

\author{Richard G. Hennig}
    \email[Correspondence email address: ]{rhennig@ufl.edu}
    \affiliation{Department of Physics, University of Florida, Gainesville, Florida 32611, USA}
    \affiliation{Department of Materials Science and Engineering, University of Florida, Gainesville, Florida 32611, USA}
    \affiliation{Quantum Theory Project, University of Florida, Gainesville, Florida 32611, USA}

\date{\today}

\begin{abstract}
To address the disconnect between computational predictions and experimental realization in AI-driven superconductor discovery pipelines, we fine-tune a diffusion foundation model on 7,183 metallic compounds with DFT-computed superconducting critical temperature ($T_c$) and predicted substitutional disorder probability, and employ classifier-free guidance to jointly steer generation toward ordered higher-$T_c$ superconductors. Relative to an unguided baseline, our joint guidance triples the fraction of unique generated structures predicted to be ordered and nearly triples their median predicted $T_c$. Structures generated from this model are then passed through a multi-stage screening pipeline that uses machine-learning models to assess disorder probability, novelty, thermodynamic and dynamical stability, and $T_c$. By comparing sets of guided, unguided and elementally substituted structures passed through the same pipeline, we find that joint guidance more than doubles the fraction of stable candidates predicted to superconduct above 5~K relative to unguided generation and nearly triples it relative to elemental substitution. Guidance, however, lowers the fraction of candidates that are thermodynamically stable. Substitution, in contrast, generates far more novel candidates and therefore yields at least as many stable, high-$T_c$ candidates. 
\end{abstract}

\maketitle

\input{sections/introduction}
\input{sections/methods}
\input{sections/results}
\input{sections/discussion}
\input{sections/conclusion}
\input{sections/acknowledgements.tex}

\section*{Data Availability}
The data and code supporting this study will be made publicly available upon journal publication.

\appendix
\input{sections/appendix1.tex}
\input{sections/appendix2.tex}

\FloatBarrier
\bibliography{sources}

\end{document}

%% file: preamble.tex
\usepackage{amsthm}
\usepackage{mathtools}
\usepackage{physics}
\usepackage{xcolor}
\usepackage{graphicx}
\usepackage[left=23mm,right=13mm,top=35mm,columnsep=15pt]{geometry} 
\usepackage{adjustbox}
\usepackage{placeins}
\usepackage[T1]{fontenc}
\usepackage{lipsum}
\usepackage{csquotes}
\usepackage{kantlipsum}
\usepackage{float}
\usepackage{dsfont}
\usepackage{placeins}

%% file: sections/introduction.tex
\section{Introduction} \label{sec:introduction}

In the inverse design of materials, one starts from a desired property and seeks the materials that exhibit it, reversing the conventional structure-to-property workflow. In recent years, generative models have made this approach practical to scale. By learning the distribution of known crystal structures, they are capable of sampling new candidates directly from a learned chemical space, reducing reliance on experimental trial and error and on costly high-throughput density functional theory (DFT) screening.

Continuous generative models, widely used for image, text, and video generation, have been adapted to crystal structures through diffusion-based~\cite{ho2020ddpm, jiao2023diffcsp, zeni2025mattergen}, flow-based~\cite{miller2024flowmm} and, more recently, stochastic-interpolant frameworks~\cite{hollmer2025omatg, prakash2026omatgrpo, hollmer2026omatgirl}. These models now underpin most inverse-design efforts, in which candidates generated from these models pass through multi-stage screening pipelines that select physically plausible structures with the desired properties.

Superconductors are a particularly attractive target given their technological potential~\cite{high_tc_applcations, prospects_of_highTc, superconducting_qubits}, but also difficult to learn since experimentally realized superconductors are scarce, and those with high critical temperatures ($T_c$) even more scarce. This relative data scarcity makes it difficult for unconditional generative models to learn the structural and chemical features that distinguish superconductors. Conditional generative models can deal with this data scarcity problem since such models can first be trained on much larger databases made up of samples which may not necessarily have the desired properties, and then be fine-tuned with respect to a sparse dataset containing structures with the desired properties~\cite{wines2023superconductors, Han_2025}. In superconductor discovery, Prakash et al.~\cite{prakash2025guideddiffusion} fine-tuned a DiffCSP foundation model, pretrained on Alexandria~\cite{schmidt2023alexandria}, an open database of theoretical DFT-relaxed inorganic compounds, on superconductors with first-principles $T_c$ labels and used classifier-free guidance on $T_c$ to propose new candidates, eighteen of which were synthesized after screening.

However, significant gaps still exist between computational predictions and experimental realization, for inverse design in general and for superconductor discovery in particular. One of the most pervasive of such gaps is due to the existence of chemical disorder, which computational predictions typically do not account for but which commonly arises upon synthesis of proposed candidates. Nearly all generative models used for inverse design are trained on databases of ordered DFT-relaxed structures, such as the Materials Project~\cite{jain2013materials} and Alexandria~\cite{Cavignac2026Alexandria}, where every site is occupied by a single element. These models therefore learn to generate crystal structures which are implicitly assumed to be ordered, rather than it being a condition that is explicitly checked. 

The problem with such a pervasive assumption is that a candidate predicted as a new ordered compound may instead form with elements randomly mixed over sites that the prediction assigns to distinct species, so that neither the predicted structure nor the predicted properties describe the real material. A prediction may not only become disordered upon synthesis, rendering computational predictions inaccurate, but may even turn out to be a disordered solid solution of a known compound, further rendering any strict claim to novelty invalid. The $T_c$-guided candidates of Ref.~\cite{prakash2025guideddiffusion} illustrate this disconnect in that although half of the eighteen synthesized compounds were superconducting above 4.2~K, none formed in their predicted structure, and the samples were instead mostly disordered body-centered cubic solid solutions or mixtures of known binary phases. The existence of disorder invalidating claims to novelty in predicted structures has been documented across the general search for new inorganic materials~\cite{leeman2024challenges, szymanski2026correction, gibson2026beenet}.

We see two natural ways to account for disorder in generative design. One is to build disorder into the generative model itself, so that it generates partially occupied structures directly~\cite{hoffmannpetersen2025disgen}. The other is to pair an established generative model with a predictive model for disorder~\cite{jakob2026disorder}. Jakob et al.\ used such a model to screen computational databases~\cite{jakob2026disorder}. Here we use it both to guide generation toward, and to restrict screening to, candidates predicted to form ordered structures. We do this by extending the $T_c$-guided model of Ref.~\cite{prakash2025guideddiffusion}, keeping its foundation model and fine-tuning data and adding predicted disorder probability as a second guidance property.

We avoid candidates predicted to be disordered in this work not because disorder is necessarily undesirable for technologically relevant superconductors, but because a structure, predicted to be ordered, which becomes disordered upon synthesis is a failed computational prediction, and every property computed from that structure, including $T_c$, is also incorrect. This is the inaccuracy we address in this work, and is one that workflows built on generative models have so far largely left unchecked.

A second question we consider in this work is what our jointly guided generative model adds over generation by elemental substitution. Generation, particularly by AI models, and screening efforts have been rightfully criticized for proposing compounds that are neither novel nor synthesizable~\cite{cheetham2024ai,leeman2024challenges}, and a recent benchmark found that ion exchange, a data-driven form of elemental substitution, yields stable new materials more often than current generative models~\cite{szymanski2025baselines}. We therefore compare the generative output of our jointly guided model against that of an unguided baseline and elemental substitution of known metals, which is the approach our previous work used to identify new superconductors~\cite{gibson2026beenet}, and pass candidates generated by all three sources through the same screening pipeline.

In this work, we extend $T_c$-guided diffusion, developed in Ref.~\cite{prakash2025guideddiffusion}, into a disorder-aware workflow that runs from generation to the selection of synthesis targets. We guide generation jointly on predicted disorder probability and $T_c$, then screen the generated structures for validity, novelty, thermodynamic and dynamical stability, and suitability for arc melting, so that the proposed candidates can be attempted with a fast and inexpensive synthesis method. Unlike many workflows that propose candidates for synthesis, including that of Ref.~\cite{prakash2025guideddiffusion}, ours performs no DFT calculations on candidates before they are proposed. Predicted stability, metallicity, dynamical stability and $T_c$ are all assessed with machine-learned models, which we accept as some loss of predictive accuracy in exchange for the ability to screen hundreds of thousands of structures at very low cost. Comparing guided generation, unguided generation and substitution at the same scale, we find that guidance substantially increases the fraction of structures predicted to be ordered and the fraction of stable candidates predicted to superconduct above 5~K, although substitution, which supplies far more novel candidates, still yields at least as many such candidates. We also show that guidance comes at a cost to thermodynamic stability, a trade-off consistent with the scarcity of stable, ordered, high-$T_c$ structures in the fine-tuning dataset.

%% file: sections/methods.tex
\section{Methods} \label{sec:methods}
\subsection{Guided Diffusion Model Architecture}

Our generative model extends the guided diffusion framework of Ref. \cite{prakash2025guideddiffusion} by first pretraining a diffusion-based model, DiffCSP \cite{jiao2023diffcsp}, on over two million crystal structures from the Alexandria database \cite{schmidt2023alexandria}, which allows the model to first learn the general features of realistic inorganic crystals regardless of any target property conditioning. We then condition this pretrained model on two scalar targets, the disorder probability predicted by the classifier of Sec.~\ref{sec:pipeline} and the superconducting critical temperature, using a smaller set of 7,183 metallic compounds with DFT-computed $T_c$ \cite{cerqueira2024sampling}.

To condition the pretrained model on our two scalar targets, we employ two separate adapter modules in the neural network's architecture. These adapter modules are applied after each message-passing layer, and node embeddings are updated according to
\begin{equation}
    \Tilde{H}_j^{(L)} = H_j^{(L)}+ \sum_{i=1}^N f^{(L)}_{\text{mixing}, i}\biggl(f^{(L)}_{\text{adapter}, i}(g_i)\biggr)\cdot \mathds{1}_i
\end{equation}
for $N$ conditioning variables. $H_j^{(L)}$ is the unconditioned embedding of node $j$ at layer $L$, $g_i$ are the sinusoidal embeddings of the conditioning signals, here the disorder probability and superconducting critical temperature, and $f_{\mathrm{adapter},i}^{(L)}$ and $f_{\mathrm{mixing},i}^{(L)}$ are learned layers. Each $\mathds{1}_i$ is a mask, drawn independently from a Bernoulli distribution during training, that switches the $i$-th conditioning signal's contribution on or off, so that the network learns to produce accurate score estimates whether or not that signal is present. During inference, each mask is instead set to 0 or 1 to give the unconditional and conditional terms of the classifier-free guidance expression.

To then sample from this fine-tuned model, we employ classifier-free guidance (CFG) \cite{ho2021classifierfree}, which has proven effective to steer generative models toward target properties \cite{zeni2025mattergen, prakash2025guideddiffusion, Jin2026}, on the joint crystal state $\boldsymbol{x}$, which combines the unconditional and conditional predictions of the fine-tuned network by
\begin{equation}
    \epsilon_{\boldsymbol{\theta}}(\boldsymbol{x}, \{c_i\}) = \epsilon_{\boldsymbol{\theta}}(\boldsymbol{x}, \varnothing)+ \sum_{i=1}^Nw_i\biggl(\epsilon_{\boldsymbol{\theta}}(\boldsymbol{x}, c_i)-\epsilon_{\boldsymbol{\theta}}(\boldsymbol{x}, \varnothing)\biggr)
    \label{eq:CFG}
\end{equation}
for $N$ conditioning variables. $\epsilon_{\boldsymbol{\theta}}(\boldsymbol{x}, \varnothing)$ is the denoised output of the unconditional model, $c_i$ are the conditioning variables, and $w_i$ are the guidance weights for each variable. Conditioning on two signals therefore results in a set of four free model hyperparameters. We discuss the choice of target values and guidance weights in Appendix \ref{app:hyperparameter_selection_appendix}.

\subsection{Generation by Elemental Substitution}
\label{sec:substitution}
To compare generation from our jointly guided model against substitution, we construct a set of structures by Wyckoff-site substitution of known metals, following Ref.~\cite{gibson2026beenet}. We queried the Materials Project for ordered, experimentally known metals with at most 12 atoms in the unit cell and an energy above the convex hull below 200~meV/atom, and kept those containing at least one of the key elements \ch{Li}, \ch{Be}, \ch{B}, \ch{C}, \ch{N}, \ch{Na}, \ch{Mg}, \ch{Si} or \ch{Ge}. In each parent, sites occupied by a key element were held fixed, and every other Wyckoff site was either left unchanged or substituted by a neighboring element in the periodic table, giving, after removing duplicates, 101,595 structures.

\subsection{Screening Pipeline Architecture}\label{sec:pipeline}
We use this joint guidance diffusion model as the generative backbone of a multi-stage pipeline to search for the most promising novel, ordered, and high-$T_c$ superconductors. We pass structures generated by unguided diffusion and elemental substitution through the same pipeline to compare the performance of our model. Central to our work, the first stage of the pipeline after generation is the substitutional disorder classifier of Jakob et al.~\cite{jakob2026disorder}, a composition-based recurrent neural network trained on about 120,000 ICSD structures measured at ambient conditions, excluding those with positional disorder. Given only a composition as an input, this model predicts the probability that the compound will form with substitutional disorder, in which two or more elements share a crystallographic site. This model was demonstrated to identify disordered compounds in held-out ICSD data with 90\% accuracy. In a separate experimental study of 34 predicted ternary compounds~\cite{gabe2026experimental}, 11 had a predicted disorder probability below 20\% and were also predicted to be thermodynamically and dynamically stable. Eight of these 11 formed in the predicted structure, whereas none of the other 23 did. We therefore remove all candidates with a predicted disorder probability above 20\%, which is a strict threshold that retains only compositions confidently predicted to be ordered.

We then apply a set of structural validity checks to remove unphysical candidates generated by the diffusion model prior to relaxation, removing any structure with a cell length outside 1–100~\AA, a minimum interatomic distance below 0.8~\AA, a minimum cell width below 1.5~\AA, more than 2000 atoms, or an atomic number outside the range supported by our relaxation model (Z = 1–99). In addition to these checks, we apply the field-standard geometric validity criterion of xtalmet~\cite{negishi2025xtalmet}, so that validity is also assessed by the same measure used in other generative-model studies. As expected, we found that these unphysical artifacts would arise more often when either conditioning signal was heavily guided to a target value underrepresented in the training data.

To efficiently relax the remaining tens of thousands of candidates, we use the OMat task head of the UMA-S-1.2 machine-learned interatomic potential \cite{UMA}, accessed via the ASE-compatible FAIRChemCalculator, allowing both atomic positions and cell dimensions to be updated. If several candidates share the same reduced composition after relaxation, we keep only the one with the lowest energy per atom, so that each composition is represented by its most stable generated polymorph. We then remove any remaining candidate whose structure matches any of those in the fine-tuning dataset, the experimentally-tagged structures of Materials Project \cite{jain2013materials}, or the same ICSD subset used in Ref.~\cite{jakob2026disorder} to encourage strict structural and compositional novelty
of our candidates.

We then apply a series of joint property screens to the remaining candidates. From each candidate's UMA-relaxed total energy, we compute its energy above the convex hull ($E_\text{hull}$), against LeMat-Bulk reference structures whose energies were computed with the same model~\cite{siron2025lematbulk,lematgenbench2025}, and its formation energy per atom ($E_f$) relative to the elements. We then predict each candidate's band gap using the multi-fidelity MEGNet model \cite{chen2019megnet, chen2021multifidelity_bandGap}, evaluated at the PBE fidelity. Candidates are screened on all three properties: we retain only candidates with $E_\text{hull} \leq 0.05$ eV/atom, $E_f < 0.02$ eV/atom, and $E_g < 0.1$ eV, discarding all others. The $E_\text{hull}$ tolerance accounts for UMA's own error on hull-energy prediction. In practice, the hull screen does almost all of the filtering; the small positive $E_f$ tolerance only allows for model error. Finally, enforcing a near-zero band gap screen ensures that remaining candidates are metals, with the $0.1$ eV tolerance added to account for the fact that MEGNet's output is a continuous regression, so observing exactly $E_g =0$ rarely occurs.

Candidates that pass these screens are then checked for dynamical stability via finite-displacement phonon calculations on the UMA-relaxed structure, using Phonopy~\cite{togo2015phonopy,togo2023phonopy}, to ensure that candidates lie at true local minima of their potential energy surface. We classify structures exhibiting imaginary modes ($\omega < -1\times10^{-3}$~THz) as dynamically unstable and remove them. Calculations were performed using the implementation released with PhononBench~\cite{han2025phononbench}.

We predict each candidate's $T_c$ with BEE-NET~\cite{gibson2026beenet} in its crystal-structure-only (CSO) mode. We use an ensemble of its 50 equivariant graph neural networks to predict the Eliashberg spectral function $\alpha^2F(\omega)$ on a 51-point grid between 0.25 and 101~meV, from which the electron--phonon coupling $\lambda$ and the frequency moments $\omega_\mathrm{log}$ and $\omega_2$ are computed, and used to obtain $T_c$ from the full Allen--Dynes equation~\cite{allen1975} with $\mu^* = 0.10$. The predicted $T_c$ in this work's pipeline, unlike in Ref.~\cite{prakash2025guideddiffusion}, is not used to screen candidates, but is used to quantify the advantage our jointly guided model offers.

To encourage the experimental realizability of our candidates, we add a final screen based on arc melting, one of the fastest and simplest routes for synthesizing intermetallic compounds, and remove all candidates unsuitable for it. We therefore exclude compositions containing noble gases, halogens, alkali metals, radioactive elements, H, O or N; any element with a boiling point below 1500~K, which would evaporate during melting (for example Mg, Zn, P and S); and the toxic elements Pb, Hg, As, Tl, Cd, Se, Sb and Te. Unlike thermodynamic and dynamical stability, which any material must satisfy to form, this screen reflects one particular synthesis route. None of the three sources was designed with this route in mind, and it also excludes well-known conventional superconductors such as \ch{MgB2} and NbN. We therefore use it only to select final candidates for synthesis, not to compare the three sources. 

%% file: sections/results.tex
\section{Results} \label{sec:results}

\subsection{\boldmath Learning Ordered and High-\texorpdfstring{$T_c$}{Tc} Superconductors}\label{sec:learning_superconductors}
Our goal is to generate the highest number of novel candidates predicted to be both ordered and superconducting at high $T_c$ from our jointly guided model. Stronger model guidance and more extreme targets push generation further into this desired region. This region is sparsely represented in the fine-tuning data, however, so guiding too hard toward it makes the model repeatedly generate the same compositions with only slight structural variations (Table~\ref{tab:guidance}). Based on the scan in Appendix~\ref{app:hyperparameter_selection_appendix}, we use guidance weights $(w_d, w_{T_c}) = (2, 2)$ with targets $T_c = 10$~K and $d = 1\%$, which are demonstrated to be capable of reaching the target region without excessive repetition.

\begin{table}
\centering
\begin{threeparttable}
\setlength{\tabcolsep}{1.5pt}
\begin{tabular}{@{}lccccc@{}}
\toprule
Guidance & $(w_{d},\, w_{T_c})$ & \makecell{Unique\\(\%)} & \makecell{Median\\$T_c$ (K)}
      & \makecell{$d < 20\%$\\(\%)} & \makecell{$T_c > 5$\,K\\\& $d < 20\%$\\(\%)} \\
\midrule
Training data          & --     & 99.1  & 0.96\tnote{a} & 23.8 & 4.0\tnote{a} \\
\midrule
None                   & (0, 0) & 89.4 & 1.49 & 18.2 & 2.4  \\
$T_c$                  & (0, 2) & 68.7 & 8.48 & 9.8  & 4.9  \\
Disorder               & (2, 0) & 72.2 & 1.04 & 68.2 & 7.8  \\
$T_c$ $\&$ disorder       & (2, 2) & 69.1 & 4.13 & 56.5 & 18.1 \\
\bottomrule
\end{tabular}
\begin{tablenotes}
\footnotesize
\item[a] Training-set $T_c$ were DFT-computed, not BEE-NET predictions.

\caption{For each set of guidance weights, 5,000 structures were generated with guidance targets fixed at $T_c = 10$~K and disorder probability $d = 1\%$, relaxed with UMA, and deduplicated with pymatgen's \texttt{StructureMatcher}, keeping the lowest-energy structure of each duplicate group. Unique is the percentage of the 5,000 generated structures that relaxed successfully and are not duplicates; the remaining columns are computed from these unique structures. $T_c$ of the generated structures is predicted with BEE-NET (CSO), and $d$ is the predicted disorder probability.}
\label{tab:guidance}
\end{tablenotes}
\end{threeparttable}
\end{table}

\begin{figure}
    \centering
    \hspace*{-5mm}\includegraphics[width=1.1\linewidth]{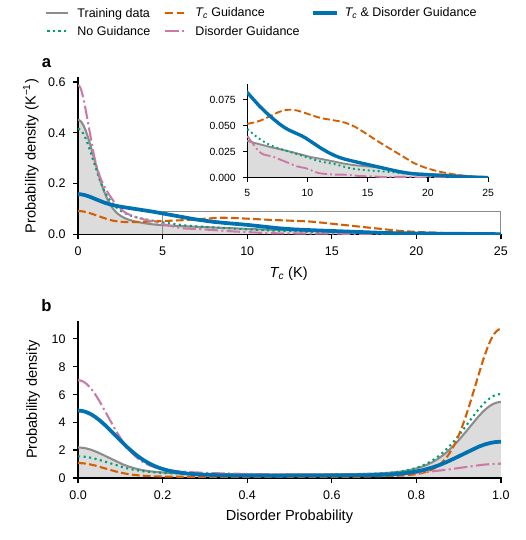}
    \caption{Kernel density estimates of predicted \textbf{a} $T_c$ and \textbf{b} disorder probability $d$ for the unique generated structures at each guidance setting in Table~\ref{tab:guidance}, with the fine-tuning data shaded in gray. $T_c$ of the generated structures is predicted with BEE-NET (CSO), while that of the fine-tuning data is computed with DFT. The $T_c$ axis is truncated at 25~K and the inset in \textbf{a} shows a magnified view of the distributions between 5 and 25~K.}
    \label{fig:kde_guidance}
\end{figure}

The data in Table~\ref{tab:guidance} and Fig.~\ref{fig:kde_guidance} compare structures generated with guidance on $T_c$ alone, on disorder alone, and on both combined, against structures generated without guidance and against the fine-tuning data. We demonstrate that guidance on a single property shifts that property in the generated candidates as expected, but works against the other. Specifically, guiding on $T_c$ alone raises the median predicted $T_c$ of unique generated structures from 1.5~K to 8.5~K, but roughly halves the fraction predicted to be ordered, from 18.2\% to 9.8\%. Similarly, guiding on disorder alone raises the ordered fraction to 68.2\% while leaving the median $T_c$ essentially unchanged. This competition is reflected in the fine-tuning dataset, in which only 15.2\% of structures with $T_c > 5$~K are predicted to be ordered, compared with 23.8\% overall, and only 17.0\% of ordered structures have $T_c > 5$~K, compared with 29.5\% of disordered ones. Joint guidance recovers a desired shift in both properties. 56.5\% of the unique structures are predicted to be ordered and their median $T_c$ rises to 4.13~K, so that 18.1\% are predicted to be both ordered and to superconduct above 5~K. This represents a 7.5 times improvement in the fraction of ordered, high-$T_c$ structures relative to the unguided model.

\subsection{Screening Results}\label{sec:screening}
\begin{figure*}[t]
  \centering
  \includegraphics[width=\textwidth]{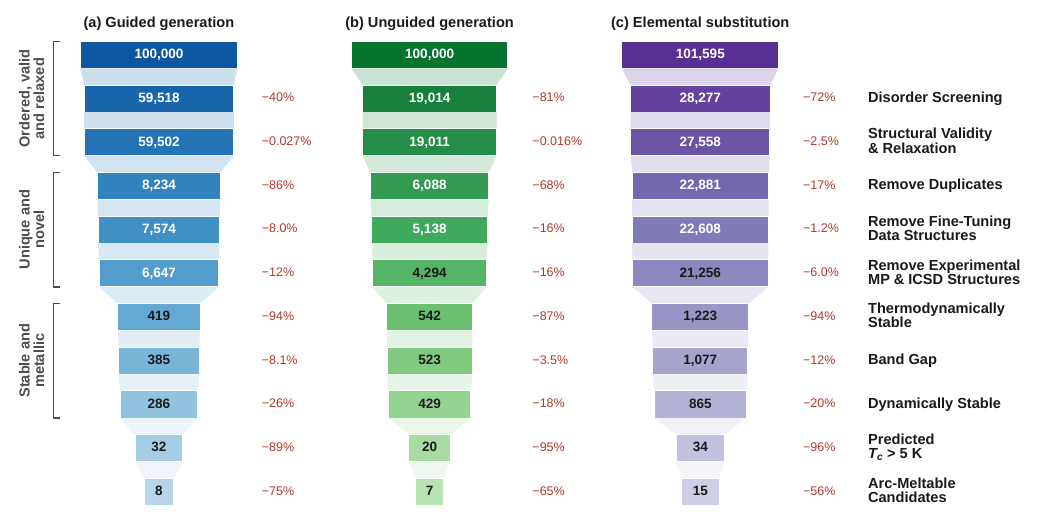}
  \caption{Screening funnels for (a) 100,000 structures generated with $T_c$ and disorder guidance, (b) 100,000 structures generated without guidance, and (c) 101,595 structures made by elemental substitution. Each row is one screening stage: the bar gives the number of candidates remaining after that stage, and the red value beside it the percentage of the previous bar's candidates that the stage removed. Duplicate removal keeps the lowest-energy structure of each reduced composition. $T_c$ is predicted with BEE-NET (CSO). Bar widths share one logarithmic scale across all panels. The three sources are compared at the predicted-$T_c$ stage; the arc-meltability screen reflects the synthesis route available to us and is shown for completeness.}
  \label{fig:funnels}
\end{figure*}

Figure~\ref{fig:funnels} follows candidates from three sources through our screening pipeline. We generate 100,000 structures with joint $T_c$ and disorder guidance, 100,000 structures by the same model with no guidance, and 101,595 structures made by elemental substitution of experimentally known metals in the Materials Project. All three sources pass through the same stages, which broadly fall into three groups: selecting ordered, structurally valid candidates that relax successfully; removing duplicates and structures already present in the fine-tuning data, the experimentally tagged Materials Project entries, or the ICSD; and retaining thermodynamically stable, metallic and dynamically stable candidates. 

The unguided run was screened only as a computational baseline, to show how guidance changes the number of candidates passing each stage. To compare how well each source produces the targeted property, we use the number of dynamically stable candidates predicted to superconduct above 5~K. The funnels also show how many candidates remain after the arc-meltability screen.

We find that the disorder screen removes most of the unguided and substituted structures but fewer than half of the guided ones, which is expected since only guided generation is steered toward compositions predicted to be ordered and replacing elements by their chemically similar periodic-table neighbors further favors mixing on the same site \cite{HumeRothery1936, Glawe2016}. The thermodynamic screen gives the largest single reduction across all three sources, retaining only 6--13\% of the novel candidates of each. Most candidates that pass the energy and band gap screens are also dynamically stable.

Joint guidance triples the fraction of generated structures that pass the disorder screen, from 19\% to 60\%, and the guided structures less often reproduce the fine-tuning data. However, guidance also increases the number of repeated compositions within a run. As a result, tripling the number of ordered structures yields only a third more distinct compositions. Guidance also halves the fraction of novel candidates within 50~meV/atom of the convex hull, from 12.6\% to 6.3\%, which is consistent with the scarcity of stable high-$T_c$ structures in the fine-tuning data (Fig.~\ref{fig:training_data}b), so that the guided run retains fewer stable candidates.

Among the dynamically stable candidates, the guided set is clearly enriched in the targeted region: 11.2\% are predicted to superconduct above 5~K, compared with 4.7\% without guidance and 3.9\% for substitution, giving 32, 20 and 34 such candidates, respectively. Many of these candidates from both generative runs contain hydrogen or technetium (19 of 32 guided and 11 of 20 unguided, compared with 4 of 34 substituted). Excluding them, guidance still raises the fraction of stable candidates predicted above 5~K to 6.7\%, from 2.6\% without guidance and 3.6\% for substitution, but substitution yields more of such candidates (30, compared with 13), because three times as many novel substituted structures enter the property screens.

The fraction of high-$T_c$ candidates is conditioned on stability, which guidance lowers, so these numbers overstate the per-sample advantage of guidance. Per 100,000 generated structures, guidance yields 32 stable candidates predicted above 5~K, which is 1.6 times as many as unguided generation. Per distinct novel composition entering the property screens, the yields of guided and unguided generation are nearly identical (0.48\% and 0.47\%, compared with 0.16\% for substitution). The stronger $T_c$ enrichment of the guided candidates is offset by their lower stability, and the advantage of guidance comes from the larger number of distinct novel compositions it carries through the disorder screen. The difference between guided and unguided generation should be read as indicative, and the arc-meltable counts below do not distinguish the two.

The final arc-meltability screen removes most of the high-$T_c$ candidates from every source, including all hydrides and technetium compounds, leaving 8, 7 and 15 arc-meltable candidates from guided generation, unguided generation and substitution, respectively. Since none of the sources were designed for this constraint, we treat these candidates as a practical shortlist for synthesis rather than as a measure of generative performance.

%% file: sections/discussion.tex
\section{Discussion} \label{sec:discussion}
Joint guidance on disorder probability and $T_c$ enriches the generated structures in ordered, high-$T_c$ candidates relative to the unguided model (Table~\ref{tab:guidance}, Fig.~\ref{fig:kde_guidance}). When the output of this model is put through the pipeline, the stability screens reduce but do not remove this enrichment, so that guided candidates that survive every screen remain more likely than unguided or substituted ones to be predicted to superconduct above 5~K. Both this enrichment and its costs trace back to the fine-tuning data (Appendix~\ref{app:training_data}), in which disorder probability and $T_c$ compete, so that few structures are both ordered and high-$T_c$. Guidance into this sparsely populated region therefore relies on the model extrapolating from few examples, and because the model was conditioned only on disorder probability and $T_c$, it had no incentive to preserve thermodynamic stability, which lowered the fraction of candidates near the convex hull. Guidance also amplifies a preference for hydrides and technetium compounds learned from the fine-tuning data, in which these compounds are more likely than others to have a $T_c$ above 5~K.

We demonstrate that elemental substitution yields at least as many stable, high-$T_c$ candidates. Few substituted structures are duplicates of known structures, so far more novel candidates reach the property screens, which outweighs its lower thermodynamic pass rate and its lower fraction of high-$T_c$ candidates. This agrees with Ref.~\cite{szymanski2025baselines}, which found that substitution more often yields stable new materials, while generative models are better at targeting specific properties.

The most important limitation of this work is that the disorder classifier uses composition only and so cannot distinguish between polymorphs. A model that predicts disorder on specific crystallographic sites, or one informed by the energetics of site mixing, would improve the prediction of whether a candidate forms as an ordered phase and would also describe the disordered structures that form instead. The same classifier also labels our fine-tuning data, steers generation and screens the generated candidates in this work, so the higher ordered fraction of guided structures reflects agreement with the classifier rather than independently confirmed order, and guidance may favor compositions on which the classifier is wrong. The classifier's agreement with synthesis outcomes in Ref.~\cite{gabe2026experimental} supports its use as a screen, but confirming that guided candidates form ordered phases requires an independent test, such as comparing the energies of ordered and site-disordered configurations, or synthesis itself.

Our workflow is also limited in scope. Guiding toward ordered structures necessarily sets aside the many known superconductors that are disordered, and generative models that represent disorder explicitly~\cite{hoffmannpetersen2025disgen,wu2026dmflow}, combined with property predictors for disordered structures, would allow such superconductors to be designed directly rather than avoided. Likewise, the only indication prior to synthesis that a candidate is superconducting comes from BEE-NET, which, like our fine-tuning data, covers only conventional, phonon-mediated superconductors, so both the generative model and the screening pipeline are limited to this class.

Classifier-free guidance can only steer generation toward label combinations represented in the fine-tuning data, and the desired ordered, stable, high-$T_c$ structures are rare there (Appendix~\ref{app:training_data}). Reinforcement learning can remove this limit since a defined reward can combine predicted disorder probability, energy above the hull, $T_c$ and synthesis constraints, and the generator is then fine-tuned on its own samples without labeled examples in the target region. We recently applied reinforcement learning to the composition channel of a crystal generator and found validated gains but also reward hacking~\cite{prakash2026omatgrpo}. The drift toward hydrides and technetium compounds seen here is a similar risk, because generation moves toward chemistries the $T_c$ predictor scores highly. Rewards that penalize ensemble uncertainty and exclude elements unsuitable for synthesis should limit this, and we are extending this approach to disorder-aware rewards.

%% file: sections/conclusion.tex
\section{Conclusion} \label{sec:conclusion}
We have presented a discovery workflow which addresses the previously unchecked assumption of structures from generative models being truly ordered in reality, which better connects computational predictions in the inverse design of materials to experimental realizability. By fine-tuning a diffusion foundation model on metallic compounds labeled with a predicted disorder probability and critical temperature, and steering generation with classifier-free guidance on both properties, we obtain candidates that are enriched in ordered, high-$T_c$ structures relative to an unguided baseline.

Passing these candidates through a multi-stage screening pipeline, alongside an unguided baseline and elemental substitution, clarifies what our generative model's guidance can and cannot achieve. We show that guidance concentrates the generated structures in the target region, more than doubling the fraction of stable candidates predicted to superconduct above 5~K relative to unguided generation and nearly tripling it relative to elemental substitution. Substitution nonetheless yields at least as many such candidates because it supplies far more novel structures. Further, guidance comes at the cost of lower thermodynamic stability and an amplified preference for hydrides and technetium compounds, which are unsuitable for arc melting.

These results point to clear routes for improvement. On the generative side, these include larger and more diverse fine-tuning data, guidance that also targets stability and excludes elements unsuitable for synthesis during sampling, reinforcement learning with rewards for order, stability and synthesizability, and feedback from synthesis outcomes into training. On the screening side these include more reliable $T_c$ predictions, literature-aware novelty checks, and structure-aware disorder models. Each of these would bring the candidates proposed by generative models closer to materials that can actually be made. 

%% file: sections/acknowledgements.tex
\section*{Acknowledgements} \label{sec:acknowledgements}
This work is supported by the U.S. National Science Foundation under
Grants No.~NSF-DMR-2118718 and No.~NSF-DMREF-2522891.
Computational resources were provided by the University of Florida HiPerGator high-performance computing system.

\section*{Author Contributions}
G.R.S., B.G., R.G.H., P.J.H.\ and J.J.H.\ conceived the project. A.F.V.\ extended the guided diffusion model to joint disorder and $T_c$ conditioning, carried out its training and the guidance analysis, built and ran the screening pipeline, and performed the elemental-substitution baseline. P.P.\ developed the original guided diffusion model. G.D.D., J.G., M.J., J.S.K., G.R.S.\ and J.J.H.\ provided the experimental perspective that shaped the screening pipeline and the properties a computational prediction should have before it is proposed for synthesis. In particular, G.D.D.\ and J.J.H.\ advised on the criteria for the arc-meltability screen. A.F.V.\ and G.D.D.\ wrote the manuscript. All authors discussed the results and revised the paper.

\section*{Use of AI Tools}
We used Claude (Anthropic; Opus 5.5 and Sonnet 5) as a coding and writing assistant. We wrote the modifications to the $T_c$-guided diffusion model ourselves and the screening pipeline scripts, then used Claude to review this code and suggest efficiency improvements. Claude also helped write the plotting scripts used to generate the figures. We wrote the manuscript text by hand and used Claude to suggest revisions. We directed each task, reviewed and tested all code and its outputs, made all scientific decisions regarding the methods, analysis and interpretation, and take full responsibility for the content of this paper.

\section*{Competing interests}
The authors declare no competing interests.

%% file: sections/appendix1.tex
\section{Selecting Model Hyperparameters}
\label{app:hyperparameter_selection_appendix}

\begin{figure*}[tp]
  \centering
  \includegraphics[width=\textwidth]{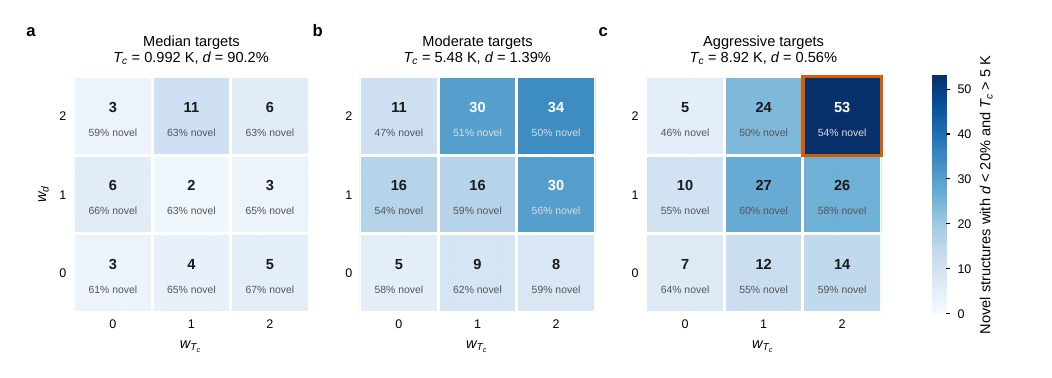}
  \caption{Each panel shows one pair of guidance targets: \textbf{a}, the median values of the fine-tuning data, and \textbf{b}, moderate and \textbf{c}, aggressive high-$T_c$, low-disorder targets. Each cell is one pair of guidance weights $(w_d, w_{T_c})$, with 500 structures generated per cell. Bold numbers give the novel structures with predicted disorder probability $d < 20\%$ and BEE-NET (CSO) $T_c > 5$~K; the text below gives the percentage of generated structures that are novel, meaning unique and absent from the fine-tuning data, experimentally tagged Materials Project entries and the ICSD.}
  \label{fig:guidance_targets}
\end{figure*}

Our joint guidance has four free parameters: a target value and a guidance weight for each of the two conditioning variables, disorder probability and $T_c$. To examine their influence on generation, we chose three pairs of targets from the distribution of the fine-tuning data (Table~\ref{tab:targets}). The median targets sit in the most populated region of the fine-tuning data, while the moderate and aggressive targets ask for increasingly ordered, high-$T_c$ structures. For each pair of targets, we generated 500 structures at every combination of weights between 0 and 2 and counted the novel structures predicted to be both ordered and superconducting above 5~K (Fig.~\ref{fig:guidance_targets}).

\begin{table}
    \centering
    \begin{threeparttable}
    \begin{tabular}{@{}lccr@{}}
    \toprule
    Targets & \makecell{Disorder\\target $d$} & \makecell{$T_c$\\target} & $n$ \\
    \midrule
    Median     & 90.2\% (median)     & 0.99\,K (median)     & 1{,}389 \\
    Moderate   & 1.39\% (10th pct.)  & 5.48\,K (75th pct.)  & 126 \\
    Aggressive & 0.56\% (5th pct.)   & 8.92\,K (85th pct.)  & 49 \\
    \midrule
    Production & 1\% (8th pct.)      & 10\,K (87th pct.)    & 59 \\
    \bottomrule
    \end{tabular}
    \end{threeparttable}
    \caption{Guidance targets used in the target scan (Fig.~\ref{fig:guidance_targets}) and in the production runs. Targets are given with their percentile in the fine-tuning set. $n$ is the number of the 7{,}183 fine-tuning structures that meet both targets, with $d \le d_\text{target}$ and $T_c \ge T_{c,\text{target}}$.}
    \label{tab:targets}
\end{table}

Without guidance, only a handful of such structures appear in each batch of 500, and guiding toward the median targets does not change this as expected. Guiding toward the moderate and aggressive targets does, and the effect is largest when both properties are strongly guided. With the aggressive targets and $(w_d, w_{T_c}) = (2, 2)$, about one in five novel structures falls in the target region, nearly ten times as often as without guidance. Guiding either disorder probability or $T_c$ on its own is not enough. A strong $T_c$ weight alone produces high predicted $T_c$ but mostly disordered structures, and a strong disorder weight alone produces ordered structures with low $T_c$, reflecting the competition between the two properties in the fine-tuning data (Appendix~\ref{app:training_data}). However, these gains come at the cost of novelty.

Since the aggressive targets with $(w_d, w_{T_c}) = (2, 2)$ produced the most novel structures that are both ordered and high-$T_c$, we used these weights for the production run, with targets close to the aggressive ones, $d = 1\%$ and $T_c = 10$~K, which are matched by a similar number of fine-tuning structures (Table~\ref{tab:targets}).

%% file: sections/appendix2.tex
\section{Training Data Limitations}
\label{app:training_data}

\begin{figure*}[tp]
  \centering
  \begin{minipage}[b]{0.49\textwidth}
    \textsf{\textbf{a}}\\
    \centering\includegraphics[height=6.5cm]{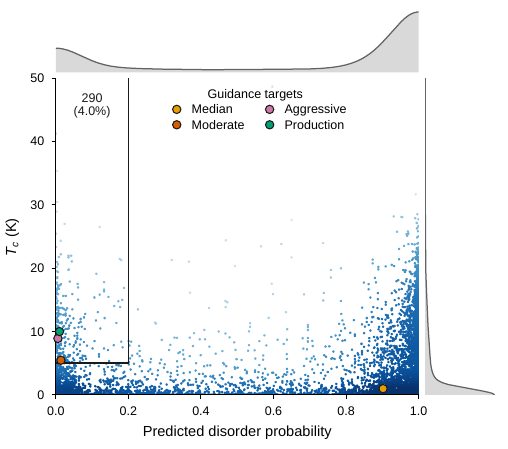}
  \end{minipage}\hfill
  \begin{minipage}[b]{0.49\textwidth}
    \textsf{\textbf{b}}\\
    \centering\includegraphics[height=6.5cm]{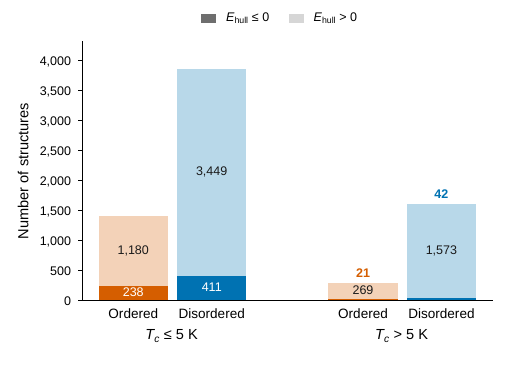}
  \end{minipage}
    \caption{\textbf{a} $T_c$ against predicted disorder probability $d$ for all 7{,}183 fine-tuning structures, including the 63 that duplicate another entry. Points are shaded by the local density of structures, with the distributions of $d$ and $T_c$ along the top and right. The outlined region contains the 290 structures (4.0\%) that are both ordered ($d < 20\%$) and superconducting above 5~K, and colored circles mark the guidance targets of Table~\ref{tab:targets}. \textbf{b} Number of fine-tuning structures in each group of $T_c$ and predicted disorder probability; the solid segment of each bar counts those on or below the convex hull ($E_\mathrm{hull} \le 0$) and the light segment the rest, with energies from relaxation with the same UMA model used for the generated candidates.}
  \label{fig:training_data}
\end{figure*}

The joint guidance is limited by two features of the fine-tuning data, the first being that ordered, high-$T_c$ structures are rare (Fig.~\ref{fig:training_data}a). Most structures with a high $T_c$ are predicted to be disordered, and only about 4\% of the fine-tuning data are both ordered and above 5~K, so the two targets pull generation in opposite directions.

The second limiting feature the joint guidance encounters from the fine-tuning data is that the few structures in the desired high $T_c$ and ordered region are rarely stable, as shown in Fig.~\ref{fig:training_data}b, which counts the fine-tuning structures lying on or below the convex hull computed with the same UMA model as for the generated candidates. Few structures of any kind sit on the hull, and the fraction drops sharply above 5~K. Among ordered structures, it falls from about one in six to about one in fourteen, leaving only 21 structures that are ordered, above 5~K and on or below the hull. Guidance toward low disorder probability and high $T_c$ therefore steers the model toward a region with very few stable examples to learn from, consistent with the lower stability of the guided candidates (Sec.~\ref{sec:screening}).